\documentclass{ws-procs9x6}

\newcommand{\siki}[1]{\eref#1}

\newcommand{\vecW}[0]{\langle W|}
\newcommand{\vecTilW}[0]{\langle \widetilde{W}|}
\newcommand{\vecV}[0]{|V\rangle }
\newcommand{\vecTilV}[0]{|\widetilde{V}\rangle }
\newcommand{\tilE}[0]{\widetilde{E}}
\newcommand{\tilD}[0]{\widetilde{D}}

\begin{document}

\title{A systematic way to find and construct exact finite dimensional matrix product stationary states}

\author{Y. Hieida$^*$}

 \address{Computer and Network Center,  Saga University ,\\
Saga 840-8502 , JAPAN\\
 $^*$E-mail:  hieida@cc.saga-u.ac.jp}

\author{T. Sasamoto}

\address{
Department of Mathematics and Informatics, \\
Faculty of Science, 
Chiba University,\\
1-33 Yayoi-cho, Inage, Chiba 263-8522, Japan }

\begin{abstract}

We explain how to construct matrix product stationary states which are composed
of finite-dimensional matrices.
Our construction explained in this article was first presented
in a part of Ref.~\refcite{0305-4470-37-42-003} for general models.  In this
article, we give more details on the treatment than in the above-mentioned
reference, for one-dimensional asymmetric simple exclusion process(ASEP).  

\end{abstract}


\bodymatter

\section{Introduction}
\label{sec:111228-1843}

Evolution of the probabilities of configurations in a class of one-dimensional
stochastic models is described by the master equation like this:
\begin{equation}
 \frac{d}{dt} \vec{P}_L(t) = -H \vec{P}_L(t) ,
\label{eq:100830-1251}
\end{equation}
where $H$ is a transition rate matrix and $\vec{P}_L(t)$ is a column vector
whose component $P(C;t)$ is the probability of finding the $L$-site system in a
configuration $C$ at time $t$.  If $\vec{P}_L(t)$ is independent
of
$t$, we obtain from Eq.~(\ref{eq:100830-1251})
\begin{equation}
 0 = H \vec{P}_L .
\label{eq:100831-1827}
\end{equation}
We call the solution $\vec{P}_L$ of this equation the \textbf{stationary state}.
Usually, as $t\to\infty$, $\vec{P}_L(t)$ reaches a stationary state $\vec{P}_L$.
So, once we have the stationary state, we can calculate, for example, the
density profile, correlations, etc. in the long-time limit.  From this
viewpoint, obtaining stationary states is important
in nonequilibrium  statistical mechanics.

For small $L$, one can solve \siki{{eq:100831-1827}} numerically and study its
properties exactly.
But, in general, 
obtaining stationary
state for \textbf{any} system size $L$ as a function of $L$ is a difficult task.

It is known that, 
for some one-dimensional stochastic models, their stationary states can be
written in a special form, namely, in the form of matrix products. We call this
the matrix product stationary state(\textbf{MPSS}).\cite{1751-8121-40-46-R01}
We consider the situation in which this matrix is independent of system size; in
this case, we can construct its stationary state for \textbf{any} system size.
%

Thus we want to find matrices which compose 
a
MPSS. 
Although, 
the dimensions of those matrices are 
generally
infinite for a model whose
interaction range is finite,\cite{0305-4470-30-9-024,Klauck1999102} 
the dimensions 
can be
 finite for some models including the ASEP under some conditions for model
parameters.\cite{0305-4470-29-13-013,0305-4470-30-13-008} We call a MPSS for the
former(resp. latter) case \textbf{an infinite(resp. a finite)-dimensional MPSS}.
To our knowledge, there is no systematic way of constructing an
infinite-dimensional MPSS.  
But, for a finite-dimensional MPSS, we found a systematic way of finding
them,\cite{0305-4470-37-42-003} which we explain in this article.

Our
construction explained in this article was first presented in a part of
Ref.~\refcite{0305-4470-37-42-003}, where the above-mentioned construction and a
systematic way of checking the validity of the obtained MPSS for an arbitrary
system size were explained for general stochastic models on the one-dimensional
lattice.
Main purpose of this article 
is to give
more details on the treatment in Ref.~\refcite{0305-4470-37-42-003}.  In order
to explain our construction concretely, we use in this article only the
one-dimensional asymmetric simple exclusion process(\textbf{ASEP}).~\cite{
SchuetzBook2001 }

It is interesting that numerical matrix product states in stochastic models (for
example, Ref.\refcite{ JPSJ.67.369 }) also play an important role in the method
of density matrix renormalization group (DMRG)~\cite{
PhysRevLett.69.2863,PhysRevB.48.10345,PeschelBook}. This article, however, does
not treat such a numerical method.

\section{Model}
\label{sec:100901-1250}

\begin{figure}
\begin{center}
\psfig{file=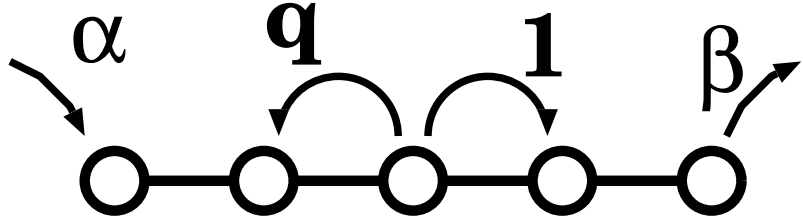}
\end{center}
\caption{
The ASEP on the one-dimensional lattice whose size is $L=5$.  A particle in the
 bulk hops to the left(resp. right) neighbor site with a rate $q$(resp. $1$) if
 the left(resp. right) neighbor site is empty.  At the leftmost site, a particle
 is injected with a rate $\alpha$ if the site is empty.  A particle at the
 rightmost site is removed with a rate $\beta$ if the site is occupied by a
 particle.
}
\label{fig:100901-0846}
\end{figure}

The ASEP in this article
\footnote{ The ASEP in this article is also called Partially ASEP(PASEP).}
is defined on the one-dimensional lattice whose size is
$L$(Fig.~\ref{fig:100901-0846}).  Each site can take two states, namely, a site
is either empty or occupied by one particle.  The time of the model is
continuous. A particle in the bulk hops to the left(resp. right) neighbor site
with a rate $q$(resp. $1$) if the left(resp. right) neighbor site is empty.  At
the leftmost site, a particle is injected with a rate $\alpha$ if the site is
empty.  A particle at the rightmost site is removed with a rate $\beta$ if the
site is occupied by a particle.

As for the master equation, for example, $\vec{P}_{L=4}(t)$ in
\siki{{eq:100830-1251}} has the following components:
\begin{equation}
\vec{P}_{L=4}(t) 
=  
\begin{pmatrix}
P(0,0,0,0;t) \\ P(0,0,0,1;t) \\ P(0,0,1,0;t) \\ P(0,0,1,1;t) \\ 
P(0,1,0,0;t) \\ P(0,1,0,1;t) \\ P(0,1,1,0;t) \\ P(0,1,1,1;t) \\ 
P(1,0,0,0;t) \\ P(1,0,0,1;t) \\ P(1,0,1,0;t) \\ P(1,0,1,1;t) \\ 
P(1,1,0,0;t) \\ P(1,1,0,1;t) \\ P(1,1,1,0;t) \\ P(1,1,1,1;t)
\end{pmatrix} .
\label{eq:100831-2232}
\end{equation}
where $\tau_k\left(1\le k\le 4\right)$ in the component $P(\tau_1, \tau_2,
\tau_3, \tau_4;t)$ of \eref{eq:100831-2232} represents the number of a particle
at the $k$-th site.
The transition rate matrix in  \siki{{eq:100830-1251}} for the ASEP 
is defined as follows:
\begin{equation}
 H:= h^{(\mathrm{L})} +
\sum_{k=1}^{L-1} h_k 
+h^{(\mathrm{R})} .
\label{eq:100830-1247}
\end{equation}
In this equation, $h^{(\mathrm{L})}$(resp. $h^{(\mathrm{R})}$) which expresses
injections of particles at the leftmost site(resp. removals of particles at the
rightmost site), is defined by
\begin{equation}
h^{(\mathrm{L})}
 :=
h^\mathrm{
left
} \otimes I^{\otimes (L-1)}
 \qquad
\text{and}
\qquad
h^{(\mathrm{R})}
 :=
I^{\otimes (L-1)}
\otimes 
h^\mathrm{
right
} \quad ,
\label{eq:100831-1109}
\end{equation}
where 
\begin{equation}
h^\mathrm{
left
}
:= 
\begin{pmatrix}
 \alpha & 0\\
 -\alpha & 0
\end{pmatrix}
\label{eq:111221-1009}
\end{equation}
\begin{equation}
h^\mathrm{
right
}
 := 
\begin{pmatrix}
0 & -\beta\\
0 & \beta
\end{pmatrix} \quad .
\label{eq:100831-2124}
\end{equation}
\siki{{eq:111221-1009}}(resp. \siki{{eq:100831-2124}} ) is  expressed in a basis
 of states whose order is $\tau_1=0,1$(resp. $\tau_L=0,1$).
In \siki{{eq:100831-1109}}, we introduce a shorthand notation of direct products
of the two-dimensional identity matrix  $I$: 
\begin{equation}
I^{\otimes n}
 := 
\overbrace{I\otimes I\otimes \cdots \otimes I}^{n} .
\end{equation}
In \siki{{eq:100830-1247}}, $h_k$, which describes hopping process between
$k$-th and $(k+1)$-th sites is defined
\begin{equation}
 h_k
:=
I^{\otimes (k-1)}
\otimes 
h_\mathrm{int}
\otimes 
I^{\otimes (L-k-1)} \quad ,
\end{equation}
where 
\begin{equation}
h_\mathrm{int}
:= 
\begin{pmatrix}
0 & 0 & 0 & 0 \\
0 & q & -1 & 0 \\
0 & -q & 1 & 0 \\
0 & 0 & 0 & 0 
\end{pmatrix} .
\label{eq:100830-1248}
\end{equation}
This is expressed in a basis of states whose order is
$(\tau_k,\tau_{k+1})=(0,0),(0,1),(1,0),(1,1)$.

\section{Matrix product stationary states(MPSSs)}
\label{sec:100901-1251}

It is possible that some models including the ASEP have a special form of a
stationary state, namely, a matrix-product stationary state(\textbf{MPSS}).  For
example, ``the vector \siki{{eq:100831-2232}} can be represented by a MPSS''
means that there exist $\langle W\mid\; ,\; E\; ,\; D$ and $\mid V\rangle$ such
that \siki{{eq:100831-2232}} is equal to
 \begin{equation}
\frac{1}{Z_{4}} 
\langle W | 
\begin{pmatrix}
E\\
D
\end{pmatrix}^{\otimes 4}
| V \rangle .
\label{eq:111229-1608}
\end{equation}
Generally, the MPSS for the ASEP is the stationary state which is equal to 
the following form
 \begin{equation}
\frac{1}{Z_{L}} 
\langle W | 
\begin{pmatrix}
E\\
D
\end{pmatrix}^{\otimes L}
| V \rangle .
\label{eq:100831-1456}
 \end{equation}
In \siki{{eq:111229-1608}} and \siki{{eq:100831-1456}},
\begin{itemize}
 \item $\langle W |$ and $|V\rangle $ are an $M$-dimensional row vector and an
       $M$-dimensional column vector, respectively. 

 \item $E$ and  $D$ are $M$-dimensional square matrices.

 \item $Z_L$ is  the normalization constant whose definition is 
\begin{equation}
Z_L
 := 
\langle W | 
\left(
E + D
\right)
^L
| V \rangle .
\label{eq:100901-0314}
\end{equation}
\end{itemize}
Hereafter, we call \siki{{eq:100831-1456}} \textbf{an $M$-dimensional MPSS}. 
It should be noted that $M$ is not always 
finite(in general, $M$ is $\infty$)
as explained in \sref{sec:111228-1843}.

In the following section \ref{sec:100831-2136}, we derive necessary conditions
for the existence of an $M(=1,2)$-dimensional MPSS for the ASEP by using its
stationary states for finite system sizes.  In the subsequent
section(\sref{sec:100831-2135}), we explain how to find $\langle W | $, $E$, $D$
and $| V \rangle $ in a two-dimensional MPSS assuming the existence of the MPSS.

\section{Necessary conditions for the existence of an $M(=1,2)$-dimensional  MPSS}
\label{sec:100831-2136}

In this section, we derive necessary conditions for the existence of an
$M(=1,2)$-dimensional MPSS for the
ASEP~\footnote{Ref.~\refcite{0305-4470-37-42-003} describes also derivation for
general $M$-dimensional MPSSs which some models with $N$ states per site have.
The ASEP in this article corresponds to $N=2$.  }.

As a tool of our derivation, we introduce a matrix form of a stationary state
$\vec{P}_{L}$.  This form has the same components as $\vec{P}_{L}$.  The
difference between the matrix form and $\vec{P}_{L}$ exists in ways of arranging
the components.  For example, the matrix form $P^{2,2}$ is defined by
rearranging components of $\vec{P}_{L=4}$(see the components of
\siki{{eq:100831-2232}} with $t$ ignored) as
\begin{equation}
 P^{2,2}
: =
\begin{pmatrix}
P(0,0,0,0) & P(0,0,0,1) & P(0,0,1,0) & P(0,0,1,1) \\ 
P(0,1,0,0) & P(0,1,0,1) & P(0,1,1,0) & P(0,1,1,1) \\ 
P(1,0,0,0) & P(1,0,0,1) & P(1,0,1,0) & P(1,0,1,1) \\ 
P(1,1,0,0) & P(1,1,0,1) & P(1,1,1,0) & P(1,1,1,1) \\ 
\end{pmatrix} .
\label{eq:100831-1637}
\end{equation}
According to this definition, the matrix form $P^{2,2}$ converted from the
MPSS(\siki{{eq:100831-1456}} with $L=4$) is
\begin{align}
 & P^{2,2}\notag\\
=
&\frac{1}{Z_4} 
\begin{pmatrix}
\vecW E E E E\vecV & \vecW E E E D\vecV & \vecW E E D E\vecV & \vecW E E D D\vecV \\ 
\vecW E D E E\vecV & \vecW E D E D\vecV & \vecW E D D E\vecV & \vecW E  D D D\vecV \\ 
\vecW D E E E\vecV & \vecW D E E D\vecV & \vecW D E D E\vecV & \vecW D E D D\vecV \\ 
\vecW D D E E\vecV & \vecW D D E D\vecV & \vecW D D D E\vecV & \vecW D D D D\vecV \\ 
\end{pmatrix} .
\label{eq:100830-1451}
\end{align}
Generally, the matrix form $ P^{m,n}$ converted from the
MPSS(\siki{{eq:100831-1456}}) can be written down as
\begin{equation}
 P^{m,n}
:=
\frac{1}{Z_{m+n}} 
\langle W | 
\begin{pmatrix}
E\\
D
\end{pmatrix}
^{\otimes m}
\begin{pmatrix}
E & D
\end{pmatrix}
^{\otimes n}
| V \rangle ,
\label{eq:100830-1516}
\end{equation}
with  $m+n = L$.

The rank of a matrix form is important to our derivation.  Suppose, by
elementary transformations, the $2^m \times 2^n $ matrix $P^{m,n}$ can be
transformed into a matrix whose form is
\begin{equation}
\left(
\begin{array}{c|c}
I_r & B\\
\hline
\multicolumn{2}{c}{O}
\end{array}
\right) ,
\label{eq:111222-0048}
\end{equation}
where $I_r$, $O$ and $B$ represents the $r\times r$ identity matrix, an
$(2^m-r)\times 2^n$ zero matrix and an $r\times (2^{n}-r)$ matrix, respectively.
Then, we can tell that the rank of $P^{m,n}$ is $r$ because elementary
transformations do not change ranks of matrices and the rank of
\siki{{eq:111222-0048}} is $r$.  Therefore, the rank of $P^{2,2}$ in
\siki{{eq:100831-1637}} is generally greater than $2$.  However, we can prove
the fact that
\begin{align}
&\text{
``if the
stationary state can be written in a two(resp. one)-dimensional}\notag\\
&\quad\text{MPSS, the rank of $P^{2,2}$ is 
at most
$2$(resp. $1$).''} 
\label{statement:111224-0010}
\end{align}
We give a  proof of this fact only for the case of a two-dimensional 
MPSS.~\footnote{This proof is slightly different from one in
Ref.~\refcite{0305-4470-37-42-003}. } 
From \siki{{eq:100830-1516}} with $m=n=2$, we obtain
\begin{equation}
Z_4
 P^{2,2}
:=
\langle W | 
\begin{pmatrix}
E\\
D
\end{pmatrix}
^{\otimes 2}
\begin{pmatrix}
E & D
\end{pmatrix}
^{\otimes 2}
| V \rangle .
\end{equation}
The right hand side of this equation can be written as a matrix product 
\begin{equation}
 A B 
\end{equation}
where
\begin{equation}
 A:= \langle W | 
\begin{pmatrix}
E\\
D
\end{pmatrix}
^{\otimes 2}
\; , \quad
 B
:=
\begin{pmatrix}
E & D
\end{pmatrix}
^{\otimes 2}
| V \rangle .
\end{equation}
$A$ is a $4\times 2$ matrix and $B$ is a $2\times 4$ matrix.
Please note that both of the rank of $A$ and the rank of $B$ is 
at most
$2$. And linear algebra tells us that the rank of $AB$ is less than or equal to
the smaller number of the rank of $A$ and the rank of $B$. This concludes the
rank of $P^{2,2}$ is not greater than $2$.

Using the fact(\ref{statement:111224-0010}), let us calculate concretely the
necessary condition of existence of an $M(=1,2)$-dimensional MPSS for
\begin{equation}
 \alpha 
>
 0\; ,\; \beta
>0
\; .
\label{eq:111223-2009}
\end{equation}
First, we obtain $\vec{P}_{L=4}$ by solving \siki{{eq:100831-1827}} with
$L=4$. And then we convert the vector form $\vec{P}_{L=4}$ into the matrix form
$P^{2,2}$.
In the following, we denote the element in the $i$-th row and the $j$-th column
of a matrix $A$ by $\left(A\right)_{i,j}$.  According to Maple, which is one of
computer algebra systems, we get
\begin{equation}
 \left(Z_4 P^{2,2}\right)_{1,1} 
= \left(\frac{\beta}{\alpha}\right)^4 ,
\label{eq:111223-2012}
\end{equation}
Because this is nonzero element(see \siki{{eq:111223-2009}}), we can multiply
the first row of $P^{2,2}$ by the inverse of \siki{{eq:111223-2012}}.  Then, we
subtract the first row of the resultant matrix multiplied by
$\left(P^{2,2}\right)_{i,1} \left(i=2,3,4\right)$ from the $i$-th row. Thus we
obtain a matrix $A^{(2)}$ like this:
\begin{equation}
\begin{pmatrix}
1 & * & * & * \\
0 & * & * & * \\
0 & * & * & * \\
0 & * & * & * \\
\end{pmatrix},
\label{eq:111223-2034}
\end{equation}
where the elements denoted by ``$*$'' mean some expressions. 
The second diagonal element of   $A^{(2)}$ is this:
\begin{align}
 \left(A^{(2)}\right)_{2,2}  
= &
 -\frac {{\beta}^{2} \left( \alpha+q \right)  \left( \alpha-1+\beta+q \right)
}{{\alpha}^{2}}
\label{eq:111224-0201}\\
\propto &\left( \alpha-1+\beta+q \right).
\label{eq:111223-2041}
\end{align}
\paragraph{case 1}

First,  we treat the case
\begin{equation}
  \alpha-1+\beta+q\ne 0.
\label{eq:111223-2038}
\end{equation}
Multiplying the second row of $A^{(2)}$ by the inverse of $
 \left(A^{(2)}\right)_{2,2} $ in \siki{{eq:111224-0201}}, we obtain the matrix
 $\widetilde{A}^{(2)}$.  Then, we subtract the second row (of
 $\widetilde{A}^{(2)}$) multiplied by $\left(A^{(2)}\right)_{i,2}\left(
 i=1,3,4\right)$ from the $i$-th row of $\widetilde{A}^{(2)}$. Thus we obtain
 the matrix, which we call $A^{(3)}$, like this:
\begin{equation}
\begin{pmatrix}
1 & 0 & * & * \\
0 & 1 & * & * \\
0 & 0 & 0 & 0 \\
0 & 0 & * & * \\
\end{pmatrix}.
\label{eq:111223-2045}
\end{equation}
It should be noted that any of the elements of the third row of this matrix is
zero.  $\left(A^{(3)}\right)_{4,3}$ is
\begin{align}
\left(A^{(3)}\right)_{4,3}
=&\frac { \left( q+1 \right)  \left( \alpha\beta+\alpha q-q+{q}^{2}+q\beta \right) 
 \left( \alpha-1+\beta+q \right) \beta}{\alpha+q}
\\
\propto &
\left( \alpha\beta+\alpha q-q+{q}^{2}+q\beta \right) 
 \left( \alpha-1+\beta+q \right)
\label{eq:111223-2209}
\end{align}
This is nonzero if we assume \siki{{eq:111223-2038}} and
\begin{equation}
\alpha\beta+\alpha q-q+{q}^{2}+q\beta
 \ne 0 .
\label{eq:111223-2210}
\end{equation}
We interchange the 3rd row and the 4th row of $A^{(3)}$ and call the resultant
matrix $A'^{(3)}$.

Because $\left(A'^{(3)}\right)_{3,3} 
=
\left(A^{(3)}\right)_{4,3} 
\ne 0$, 
we can multiply the 3rd row of $A'^{(3)}$ by the inverse of $
\left(A'^{(3)}\right)_{3,3}$.  We call the obtained matrix
$\widetilde{A'}^{(3)}$.  Then, we subtract the 3rd row (of
$\widetilde{A'}^{(3)}$) multiplied by $\left(A'^{(3)}\right)_{i,3}$ from the
$i(=1,2,4)$-th row. Thus we obtain the matrix
 \begin{equation}
A^{(4)}
:=
  \left[ \begin {array}{cccc} 1&0&0&0\\ \noalign{\medskip}0&1&0&-{
\frac {\beta+q}{\beta}}\\ \noalign{\medskip}0&0&1&{\beta}^{-1}
\\ \noalign{\medskip}0&0&0&0\end {array} \right] 
 \end{equation}
Comparing this with \siki{{eq:111222-0048}}, we know the rank of $A^{(4)}$ is 3.

\paragraph{case 2}

We go back to \siki{{eq:111223-2209}} and treat the case where
\siki{{eq:111223-2210}} does not hold. In this case, 
\begin{equation}
\alpha\beta+\alpha q-q+{q}^{2}+q\beta
=
0.
\end{equation}
We solve this with respect to $\alpha$ and obtain
\begin{equation}
\alpha
=
-{\frac {q \left( -1+\beta+q \right) }{\beta+q}}.
\end{equation}
Substituting this into $A^{(3)}$, we obtain
\begin{equation}
\left[ \begin {array}{cccc} 1&0&-{\frac {q \left( -1+\beta+q \right) 
^{2}}{\beta\, \left( \beta+q \right) }}&-{\frac {q \left( -1+\beta+q
 \right) ^{2}}{{\beta}^{2} \left( \beta+q \right) }}
\\ \noalign{\medskip}0&1&-{\frac {q\beta-\beta-2\,q+{q}^{2}}{\beta+q}}
&-{\frac {2\,{q}^{2}+3\,q\beta-2\,q-\beta+{\beta}^{2}}{\beta\, \left( 
\beta+q \right) }}\\ \noalign{\medskip}0&0&0&0\\ \noalign{\medskip}0&0
&0&0\end {array} \right] .
\label{eq:111223-2307}
\end{equation}
Comparing this with \siki{{eq:111222-0048}}, we know the rank of
\siki{{eq:111223-2307}} is $2$.

\paragraph{case 3}

We go back to \siki{{eq:111223-2041}} and treat the case where
\siki{{eq:111223-2038}} does not hold. In this case, 
\begin{equation}
  \alpha-1+\beta+q
=
0.
\end{equation}
We solve this with respect to $\alpha$ and obtain
\begin{equation}
\alpha
=
1-\beta-q
\end{equation}
Substituting this into $A^{(2)}$, we obtain
\begin{equation}
\left[ \begin {array}{cccc} 1&-{\frac {-1+\beta+q}{\beta}}&-{\frac {-
1+\beta+q}{\beta}}&{\frac { \left( -1+\beta+q \right) ^{2}}{{\beta}^{2
}}}\\ \noalign{\medskip}0&0&0&0\\ \noalign{\medskip}0&0&0&0
\\ \noalign{\medskip}0&0&0&0\end {array} \right] 
\label{eq:111225-1840}
\end{equation}
Comparing this with \siki{{eq:111222-0048}}, we know the rank of
\siki{{eq:111225-1840}} is $1$.

In summary, 
\begin{itemize}
 \item  the rank of $P^{2,2}$ is 3 if $\alpha-1+\beta+q\ne 0$ and
        $\alpha\beta+\alpha q-q+{q}^{2}+q\beta \ne 0$ are satisfied.
\footnote{
The reason why the rank is not 4 in this case is that 
the ASEP has infinite-dimensional MPSSs with 
\begin{equation}
 (DE - q ED) \propto (E+D)\; ,\quad
D\mid V\rangle \propto \mid V\rangle
\label{eq:111228-0501}
\end{equation}
and thus $EE\mid V\rangle$,$ED\mid V\rangle$,$DE\mid V\rangle$ and $DD\mid
        V\rangle$ can be written as linear combinations of 
        (at most)
        three independent
        vectors. 
}
 \item the rank of $P^{2,2}$ is 2 if $\alpha-1+\beta+q\ne 0$ and
\begin{equation}
 \alpha\beta+\alpha q-q+{q}^{2}+q\beta = 0
\label{eq:111224-0020}
\end{equation}
are satisfied.
 \item the rank of $P^{2,2}$ is 1 if
\begin{equation}
 \alpha-1+\beta+q= 0
\label{eq:111224-0021}
\end{equation}
 is satisfied.
\end{itemize}

So we can conclude that according to the fact(\ref{statement:111224-0010}),
\siki{{eq:111224-0020}}(resp. \siki{{eq:111224-0021}}) is the necessary
condition of existence of a two(resp. one)-dimensional MPSS.  These conditions
agree with the known
result.\cite{0305-4470-29-13-013,0305-4470-30-13-008,0305-4470-32-41-306,JPSJ.69.1055}.

\section{ How to find two-dimensional matrices and vectors in MPSSs for the ASEP}
\label{sec:100831-2135}

For the ASEP, the MPSS is constructed according to \siki{{eq:100831-1456}} with
matrices($E$ and $D$) and vectors($\vecW$ and $\vecV$). Hereafter we
call the matrices and vectors in the MPSS ``\textbf{the set of matrices}''.

In this section, we explain our way of finding the set of matrices in the
two-dimensional MPSS for the ASEP when $\vec{P}_L$ can be expressed as a MPSS.
It should be noted that our way of finding is not restricted to two-dimensional
MPSS for the ASEP. Our way can be applied to finite $M(\ge 2)$-dimensional MPSSs
for models which have the finite numbers 
$N(\le M)$
 of states per
site~\cite{0305-4470-37-42-003}.

A similarity transformation plays a crucial role in our way of finding.  So we
begin with an explanation of the transformation in the MPSS.

Let us consider a similarity transformation for the set of matrices
\begin{equation}
 \widetilde{E} := S^{-1} E S\; , \quad 
\widetilde{D} := S^{-1} D S\; , \quad
\langle \widetilde{W} | := \langle W | S\; , \quad
| \widetilde{V}\rangle := S^{-1} |V \rangle ,
\label{eq:100830-1514}
\end{equation}
where $S$ is a $2\times 2$ matrix.  Please note that, in this section, $E$ and
$D$ are $2\times 2$ matrices and $\langle W |$ and $|V \rangle$ are
two-dimensional vectors.  The transformation \siki{{eq:100830-1514}} does not
change a matrix form Eq.~(\ref{eq:100830-1516}) including, for example,
$P^{2,2}$ in \siki{{eq:100830-1451}}.

We make a choice $S$ in \siki{{eq:100830-1514}} as
\footnote{ 
The following set of
Eqs. (\ref{eq:111225-2228-1}) and (\ref{eq:111225-2228-2})
is what  the equation
\begin{equation}
 S := 
\begin{pmatrix}
E|V\rangle 
&
D|V\rangle 
\end{pmatrix} ,
\label{eq:100901-0322}
\end{equation}
which is Eq. (2.28) in Ref.~\refcite{0305-4470-37-42-003}, means. 
}
\begin{equation}
 S := 
\begin{pmatrix}
\kappa & \mu\\
\lambda & \nu
\end{pmatrix} ,
\label{eq:111225-2228-1}
\end{equation}
where
\begin{equation}
\begin{pmatrix}
\kappa \\
\lambda
\end{pmatrix}
:= E\mid V\rangle \; ,\quad
\begin{pmatrix}
\mu\\
\nu
\end{pmatrix}
:= D\mid V\rangle .
\label{eq:111225-2228-2}
\end{equation}
Our choice, namely, the set of Eqs. (\ref{eq:111225-2228-1}) and
(\ref{eq:111225-2228-2}), plays another crucial role in our way of finding.
Because of $S^{-1} S = 
\begin{pmatrix}
1&0\\
0&1
\end{pmatrix}
$,
\begin{equation}
 S^{-1}
 \begin{pmatrix}
\kappa\\
\lambda
\end{pmatrix}
=
 \begin{pmatrix}
1\\
0
\end{pmatrix}\; ,\quad
 S^{-1}
 \begin{pmatrix}
\mu\\
\nu
\end{pmatrix}
=
 \begin{pmatrix}
0\\
1
\end{pmatrix}.
\label{eq:111225-2324}
\end{equation}
Using Eqs.  (\ref{eq:100830-1514}) and (\ref{eq:111225-2228-2}) and
(\ref{eq:111225-2324}), we can derive
\footnote{
The following set of Eqs.~(\ref{eq:100830-1544}) and (\ref{eq:100830-1546}) is
what the equation 
\begin{equation}
  \begin{pmatrix}
   1 & 0\\
   0 & 1
  \end{pmatrix}
=
\begin{pmatrix}
 \widetilde{E}|\widetilde{V}\rangle 
&
\widetilde{D}|\widetilde{V}\rangle
\end{pmatrix},
\label{eq:100830-1549}
 \end{equation}
which is Eq. (2.29) in Ref.~\refcite{0305-4470-37-42-003}, means.
}
\begin{equation}
\widetilde{E}|\widetilde{V}\rangle 
 =  S^{-1} E|V\rangle 
=   S^{-1} 
 \begin{pmatrix}
\kappa\\
\lambda
\end{pmatrix}
 =
 \begin{pmatrix}
1\\
0
\end{pmatrix}.
\end{equation}
\begin{equation}
\therefore
\widetilde{E}|\widetilde{V}\rangle 
=
 \begin{pmatrix}
1\\
0
\end{pmatrix}.
\label{eq:100830-1544}
\end{equation}
Similarly, we can derive also the equation
\begin{equation}
\widetilde{D}|\widetilde{V}\rangle  
=
 \begin{pmatrix}
 0\\
1
\end{pmatrix}.
\label{eq:100830-1546}
\end{equation}

Using Eqs.~(\ref{eq:100830-1544}) and (\ref{eq:100830-1546}), we can derive the
equations for $\tilE$ and $\tilD$
\footnote{
The following
set of Eqs.~(\ref{eq:111226-0119}) and (\ref{eq:111226-0225}) is what the
equation 
\begin{equation}
Z_3 P^{1,2}
=
Z_2 P^{1,1}
\begin{pmatrix}
\tilE & \tilD
\end{pmatrix},
\label{eq:100830-1735}
\end{equation}
which is Eq. (2.35) in Ref.~\refcite{0305-4470-37-42-003}, means.
}
\begin{equation}
Z_3\; P^{1,2} [1:2, 1:2]
= 
\left(Z_2 P^{1,1}\right) \tilE
\label{eq:111226-0119}
\end{equation}
and
\begin{equation}
Z_3\; P^{1,2} [1:2, 3:4]
= 
\left(Z_2 P^{1,1}\right) \tilD .
\label{eq:111226-0225}
\end{equation}
In Eqs.~(\ref{eq:111226-0119}) and (\ref{eq:111226-0225}), we introduce the
notation $A[b:c,d:e]$ for a submatrix of a matrix $A$, which is constructed by
selecting the row range from the $b$-th row to the $c$-th row and the column
range from the $d$-th column and $e$-th column.

The derivation of Eqs.~(\ref{eq:111226-0119}) and (\ref{eq:111226-0225})
is postponed in the end of this section.
\footnote{
The derivation starts at the paragraph containing \siki{{eq:111226-0450}}.
}

Obtaining $\tilE$ and $\tilD$ from Eqs.~(\ref{eq:111226-0119}) and
(\ref{eq:111226-0225}) under \siki{{eq:111224-0020}}
\footnote{
In this article, we assume that the necessary condition \siki{{eq:111224-0020}} for
existence of a two-dimensional MPSS is also sufficient condition.
}
is, of course, easy: By solving \siki{{eq:100831-1827}} for $L=2$ under
\siki{{eq:111224-0020}}, we obtain $\vec{P}_{L=2}$. Then we convert the matrix
form $Z_2\; P^{1,1}$ from $Z_2\; \vec{P}_{L=2}$ as we converted
\siki{{eq:100831-1637}} from $\vec{P}_{L=4}$. Similarly, we obtain the matrix
form $Z_3\; P^{1,2}$.
By multiplying the inverse of the $2\times 2$ matrix $Z_2 P^{1,1}$ from the left
of both sides of \siki{{eq:111226-0119}}(resp. \siki{{eq:111226-0225}}), we can
obtain the solution $\tilE$(resp. $\tilD$).

Now that we know $\tilE$ and $\tilD$, we can obtain $\vecTilV$(resp. $\vecTilW$)
from \siki{{eq:100830-1544}} or \siki{{eq:100830-1546}}(resp. the set of
Eqs.~(\ref{eq:111226-0154}) and (\ref{eq:111226-0143}) or the set of
Eqs.~(\ref{eq:111226-0154}) and (\ref{eq:111226-0155}).
We can show~\cite{0305-4470-37-42-003} that the thus obtained $\tilE$, $\tilD$,
$\vecTilV$ and $\vecTilW$ are equivalent to the known
results.\cite{0305-4470-29-13-013,0305-4470-30-13-008,0305-4470-32-41-306,JPSJ.69.1055}

In the remaining part of this section, we derive only \siki{{eq:111226-0119}}(
we can derive \siki{{eq:111226-0225}} in the similar way).

By the definition \siki{{eq:100830-1516}} and the similarity
transformation \siki{{eq:100830-1514}}, we have
\begin{equation}
Z_2 P^{1,1}
=
\vecTilW
\begin{pmatrix}
\tilE\\
\tilD
\end{pmatrix}
\begin{pmatrix}
\tilE & \tilD
\end{pmatrix}
\vecTilV .
\label{eq:111226-0450}
\end{equation}
Therefore
\begin{equation}
Z_2 P^{1,1}
=
\begin{pmatrix}
\vecTilW\tilE^2 \vecTilV &\vecTilW \tilE\tilD \vecTilV\\
\vecTilW\tilD\tilE \vecTilV &\vecTilW\tilD^2\vecTilV
\end{pmatrix}.
\label{eq:111226-0139}
\end{equation}
Using 
\begin{equation}
\vecTilW\tilE 
=: \left(x_1 , y_1\right) ,
\label{eq:111226-0143}
\end{equation}
$\vecTilW\tilE^2 \vecTilV $ in \siki{{eq:111226-0139}} is equal to $x_1$ in
\siki{{eq:111226-0143}} because  
\begin{equation}
\vecTilW\tilE^2 \vecTilV
=  \vecTilW\tilE 
\begin{pmatrix}
1\\
0
\end{pmatrix}
= x_1,
\label{eq:111226-0141}
\end{equation}
where the first (resp. second) equal sign in \siki{{eq:111226-0141}} holds
because of \siki{{eq:100830-1544}} (resp. \siki{{eq:111226-0143}}).  
In the similar manner, using also
\siki{{eq:100830-1546}}, we can show
\footnote{
The set of Eqs.~(\ref{eq:111226-0154}), (\ref{eq:111226-0143})
and(\ref{eq:111226-0155}) is what the equation 
\begin{equation}
Z_2 P^{1,1}
=
\begin{pmatrix}
\vecTilW\tilE\\
\vecTilW\tilD
\end{pmatrix}
\end{equation}
which is Eq. (2.30) in Ref.~\refcite{0305-4470-37-42-003}, means.
}
\begin{equation}
 Z_2 P^{1,1}
=
\begin{pmatrix}
 x_1&y_1\\
 x_2&y_2
\end{pmatrix},
\label{eq:111226-0154}
\end{equation}
where we introduced
\begin{equation}
 \vecTilW\tilD 
=: \left(x_2 , y_2\right) .
\label{eq:111226-0155}
\end{equation}
By the same way as the derivation of \siki{{eq:111226-0450}}, we have
\begin{equation}
Z_3 P^{1,2}
=
\vecTilW 
\begin{pmatrix}
\tilE\\
\tilD
\end{pmatrix}
\begin{pmatrix}
\tilE 
&
 \tilD
\end{pmatrix}
^{\otimes 2}
\vecTilV .
\end{equation}
From this equation,
 we obtain the left part of $Z_3 P^{1,2}$
\begin{equation}
Z_3 P^{1,2}[1:2, 1:2]
=
\begin{pmatrix}
\vecTilW\tilE^3 \vecTilV & \vecTilW\tilE^2 \tilD \vecTilV\\
\vecTilW\tilD\tilE^2\vecTilV & \vecTilW\tilD\tilE\tilD \vecTilV
\end{pmatrix}.
\label{eq:111226-0211}
\end{equation}
Using \siki{{eq:100830-1544}},
the left upper element in the right 
hand
side of this equation can be transformed as
\begin{equation}
\vecTilW\tilE^3 \vecTilV 
=
\left(\vecTilW\tilE^2 \right)
\begin{pmatrix}
 1\\
0
\end{pmatrix}.
\label{eq:111226-0213-1}
\end{equation}
Using \siki{{eq:100830-1546}}, the right upper element in the right hand side of
\siki{{eq:111226-0211}} can be transformed as
\begin{equation}
 \vecTilW\tilE^2 \tilD \vecTilV
=
\left(\vecTilW\tilE^2 \right)
\begin{pmatrix}
0\\
1
\end{pmatrix}.
\label{eq:111226-0213-2}
\end{equation}
Eqs.~(\ref{eq:111226-0213-1}) and (\ref{eq:111226-0213-2}) mean that components
of the first row of\\$Z_3 P^{1,2}[1:2, 1:2]$ is equal to components of
$\vecTilW\tilE^2\left( = \left(x_1 , y_1\right) \tilE\right)$.  Similarly, we
can show that components of the second row of $Z_3 P^{1,2}[1:2, 1:2]$ is equal
to components of $\left(x_2 , y_2\right) \tilE$. Therefore we obtain
\begin{equation}
Z_3 P^{1,2}[1:2, 1:2]
=
\begin{pmatrix}
\left( x_1, y_1\right)\tilE\\
\left(x_2, y_2\right)\tilE
\end{pmatrix},
\end{equation}
that is,
\begin{equation}
Z_3 P^{1,2}[1:2, 1:2]
=
\begin{pmatrix}
 x_1&y_1\\
 x_2&y_2\\
\end{pmatrix} 
\tilE .
\label{eq:111226-0223}
\end{equation}
This equation and \siki{{eq:111226-0154}} means \siki{{eq:111226-0119}}.
In the similar way, we can show \siki{{eq:111226-0225}}.

\section{Summary}

In this article, after a brief introduction, we described the asymmetric simple
exclusion process (ASEP) in Sec.~\ref{sec:100901-1250} and a matrix product
stationary state(MPSS) for the ASEP in Sec.~\ref{sec:100901-1251}.

Furthermore,
 we have explained:
\begin{itemize}
 \item [(i)] a systematic way to find necessary conditions for the existence of an
$M(=1,2)$-dimensional MPSS(\siki{{eq:100831-1456}}) in
Sec.~\ref{sec:100831-2136};

 \item [(ii)] a systematic way by which the two-dimensional matrices($E$ and $D$) and
       vectors($\vecW$ and $\vecV$) in the MPSS(\siki{{eq:100831-1456}}) can be
       constructed from the stationary states for the two-site system and the
       three-site system if the condition in (i) is also a sufficient condition.

\end{itemize}

The method (i)(resp. (ii)) is applicable to $M(\ge 2)$-dimensional
MPSSs(\siki{{eq:100831-1456}}) for not only the ASEP but also models which have
$N(\ge 2)$(resp. $N(\le M)$) states per site.\cite{0305-4470-37-42-003}

For other examples and details, please see
Ref.~\refcite{0305-4470-37-42-003}. This reference explains also a systematic
way to check the validity of the obtained MPSS for arbitrary system sizes in the
restricted models.

\section*{Acknowledgments}

The authors would like to thank Tomotoshi Nishino for informing us of the
workshop, where partial contents written in this article was presented.

\bibliographystyle{ws-procs9x6}
\bibliography{100830-1046--proceedings-of-MatrixProduct-workshop}

\end{document}